\documentclass[conference]{IEEEtran}
\ifCLASSINFOpdf
\usepackage[pdftex]{graphicx}
\graphicspath{{./pdf/}}
\else
\usepackage[dvips]{graphicx}
\graphicspath{{./pdf/}}
\fi
\begin{document}
	%
	\title{Data Cache Prefetching with Perceptron Learning}

	\author{\IEEEauthorblockN{Haoyuan Wang}
		\IEEEauthorblockA{School of Automation\\
			Huazhong University of Science and Technology\\
			Wuhan, China\\
			Email: lkwywhy@hust.edu.cn}
		\and
		
		\IEEEauthorblockN{Zhiwei Luo}
		\IEEEauthorblockA{School of Automation\\
			Huazhong University of Science and Technology\\
			Wuhan, China\\
			Email: zhiweiluo@hust.edu.cn}
	
	}
	
	
	%


	\maketitle
	
	\begin{abstract}
		Cache prefetcher greatly eliminates compulsory cache misses, by fetching data from slower memory to faster cache before it is actually required by processors. Sophisticated prefetchers predict  next use cache line by  repeating program's historical spatial and temporal memory access pattern. However, they are error prone and the mis-predictions lead to cache pollution and exert extra pressure on memory subsystem. In this paper, a novel scheme of data cache prefetching with perceptron learning is proposed. The key idea is a two-level prefetching mechanism. A primary decision is made by utilizing previous table-based prefetching mechanism, e.g. stride prefetching or Markov prefetching, and then, a neural network, perceptron is taken to detect and trace program memory access patterns, to help reject those unnecessary prefetching decisions. The perceptron can learn from both local and global history in time and space, and can be easily implemented by hardware. This mechanism boost execution performance by ideally mitigating cache pollution and eliminating redundant memory request issued by prefetcher. Detailed evaluation and analysis were conducted based on SPEC CPU 2006 benchmarks. The simulation results show that generally the proposed scheme yields a geometric mean of 60.64\%-83.84\% decrease in prefetching memory requests without loss in instruction per cycle(IPC)(floating between -2.22\% and 2.55\%) and cache hit rate(floating between -1.67\% and 2.46\%). Though it is possible that perceptron may refuse useful blocks and thus cause minor raise in cache miss rate, lower memory request count can decrease average memory access latency, which compensate for the loss, and in the meantime, enhance overall performance in multi-programmed workloads.
	\end{abstract}
	

	%
	\IEEEpeerreviewmaketitle

	\section{Introduction}
	
	Memory wall has become increasingly a bottleneck for processor performance boost\cite{refStupid, refMemoryWall}. Prefetching is a useful technique for addressing the memory wall issue for its unique advantage of not incurring large area and energy penalty\cite{refEffective}. It effectively hide the gap between memory access latency and cache access latency. Ideally, a prefetcher can eliminate nearly all the cache misses, thus help achieve the execution performance close to that of a perfect cache\cite{refPerfect, refPerfect2}.  Due to these advantages, prefetching is now being widely used in high-performance processors, for example, Intel Xeon\cite{refHaswell, refIntel} and IBM POWER\cite{refIBMPower6, refIBMPower8} .
	
	However, prefetching is not panacea\cite{refSurvey}. In processor, prefetcher predicts 'next use' cache line and place it before required by the processor. Due to spatial and temporal predictability of memory access patterns\cite{refSpatial, refTemporal}, sophisticated prefetching methods makes prediction by tracing the route to detect fixed stride between two consecutive memory references. It's indeed useful for regular data accesses, but error prone when facing irregular ones. Naive prefetching is harmful because it evicts the useful cache lines and  brings useless cache lines, which leads to cache space consuming and performance degrading\cite{refFeedback}, also known as cache pollution.
	
	In order to ameliorate the problem, an operative way is to let prefetching learn from and adapt to the memory access pattern of the instruction stream. In this paper, a novel scheme of two-level prefetcher is proposed, which combine traditional methods(stride prefetching\cite{refStride1, refStride2, refStride3, refStride4, refStride5}, Markov prefetching\cite{refMarkov}, etc.) with perceptron learning. The first level provides suggestion and necessary information, and the second level, our perceptron, will make final decision, according to the dynamically detected memory reference patterns.
	
	According to our simulation, this two-level prefetcher can bring significant cache level optimizations. Compared with traditional methods(stride prefetching and Markov prefetching), it cuts down a geometric mean of 60.64\% and 83.84\%, respectively, unnecessary memory request issued by prefetcher. Consequently, the pressure on next level cache or main memory largely mitigates due to  the large decrease of unnecessary prefetching requests. Meanwhile, perceptron does not exert negative influence on hit rate (floating between -1.67\% and 2.46\%) and instruction per cycle(IPC)(floating between -2.22\% and 2.55\%).
	
	The key idea of our perceptron learning derives from Rosenblatt perceptron\cite{refPerceptron1, refPerceptron2}, the simplest-structured artificial neural network, but just one neuron is used. The neuron  has good learning ability for linear pattern classification task. The inputs of our perceptron are quantified features containing  local and global, time and space history of instruction stream. The weights show highly sensitivity to each program's historical memory references behavior. In original method, the output of the perceptron is the dot product of the weights and a vector of inputs.  In  our work, the sum of input and weights product are computed, and the decision will be determined by the sign of result.
	
	This paper is organized into seven sections. Section 2 introduces the related work of sophisticated prefetching schemes and perceptron learning in micro-architecture. Section 3 gives the detailed describe of perceptron learning applied in prefetching and depicts the systematic overview of our two-level prefetching scheme. Section 4 discussed the methodology of the implementation and overhead of the scheme along with the configurations and workloads in our simulation. Section 5 shows the results of simulation with analysis and explanations. Section 6 draws the conclusion of this paper, and states the future work.

	\section{Related Work}
	
	Prefetching is performed in either hardware or software. We focus on hardware-based prefetching for its unique advantage that it does not 'steal' cycles from the execution of instruction stream\cite{refSurvey}. With the association of extra storage, memory access behavior is recorded, and thus patterns are detected from those historical information, then data that is expected to be referenced soon are prefetched. According to the storage structure, hardware-based prefetchers are mainly classified into table-based prefetching, tag-based prefetching and helper-thread based prefetching\cite{refSurvey}. As mentioned above, quantified features extracted from historical memory access behavior consist the input of our perceptron. Thus, a storage structure that provides instant and well-organized information for next level prefetcher will be the candidate for the first level prefetching. In our scheme, table-based prefetchers, stride prefetching\cite{refStride1, refStride2, refStride3, refStride4, refStride5} and Markov prefetching\cite{refMarkov} with global history buffer(GHB)\cite{refGHB1,refGHB2} are chosen.  Furthermore, prefetcher should be strict with timeliness, because a prefetching response with off-limit latency will be meaningless. In order to catch up with processor's clock speed, prefetching determination should be given in seldom clock cycles (varying from cache level).
	Consequently, artificial neural networks with complex hardware implementation are infeasible for applying in prefetching. Noticing that perceptron learning enjoys the merit of costing only one easily hardware-implemented neuron to achieve good learning ability for linear classification task, it is chosen as the voter in the second level prefetching.
	
	In this section, we will review those data prefetching methods that prefetch based on history information recorded in the table. Also, perceptron learning's applying in micro-architecture is also mentioned.

	\textbf{a) Stride Prefetching: }
	Stride prefetcher is a widely used prefetcher\cite{refStride1, refStride2, refStride3, refStride4, refStride5}. Conventional stride prefetcher use a table to save stride-related information and find out stride pattern relative to current cache miss address, for example, $A$. Once the stride is recognized and confirmed according to the memory access behavior recorded in the table, prefetching request about address $A + s$, $A + 2s$, ..., $A + ds$ will be issued right after prefetcher is triggered -- where $s$ is the detected stride and $d$ is the prefetch degree, an implementation dependent prefetch look-ahead distance. In more aggressive prefetch implementation, a higher value will be used for $d$. In high-performance processors, stride prefetching is the most widely used one\cite{refIBMPower6,refIBMPower8}.
	
	\textbf{b) Markov Prefetching: }
	The Markov model can server as a basis for global correlation prefetching mechanisms\cite{Markov}.  It works on the basis of an assumption that the miss address stream can be approximated by a Markov diagram, a probabilistic state machine. This diagram keeps records on the probability of transition from one miss address to another miss address. The nodes in the diagram represent each cache misses, and the arcs means every transition probability from one node to another that the arc connects. Assuming that the execution pattern is repetitive, the Markov diagram are used to predict the following next possible cache miss address. Based on Markov diagram, prefetching algorithms can use information from nodes and arcs in a number of ways to predict future addresses.
	
	\textbf{c) Global History Buffer: }
	Global history buffer\cite{refGHB1,refGHB2} is a FIFO structure that keeps recent cache miss addresses, which can be used to implement a number of different prefetch method, including stride prefetch and Markov prefetch. What's more, the GHB provides a complete cache miss history, both local and global, time and space information can be extracted from the sequential GHB entries.
	The instant and comprehensive information extracted from GHB will enable perceptron to dynamically trace program memory access pattern can make reasonable determinations.
	
	\textbf{d) Perceptron learning in micro-architecture: }
	Perceptron learning has already been successfully used in reuse prediction\cite{refDalao} and branch prediction\cite{refBranch}. It has been proved that perceptron learning allows finding independent correlations between multiple features, and the application gives rise to an optimization of performance.

	\section{The Main Idea}
	\begin{center}
		\begin{figure}[ht]
			\hspace{-2em}
			\includegraphics[scale=0.59]{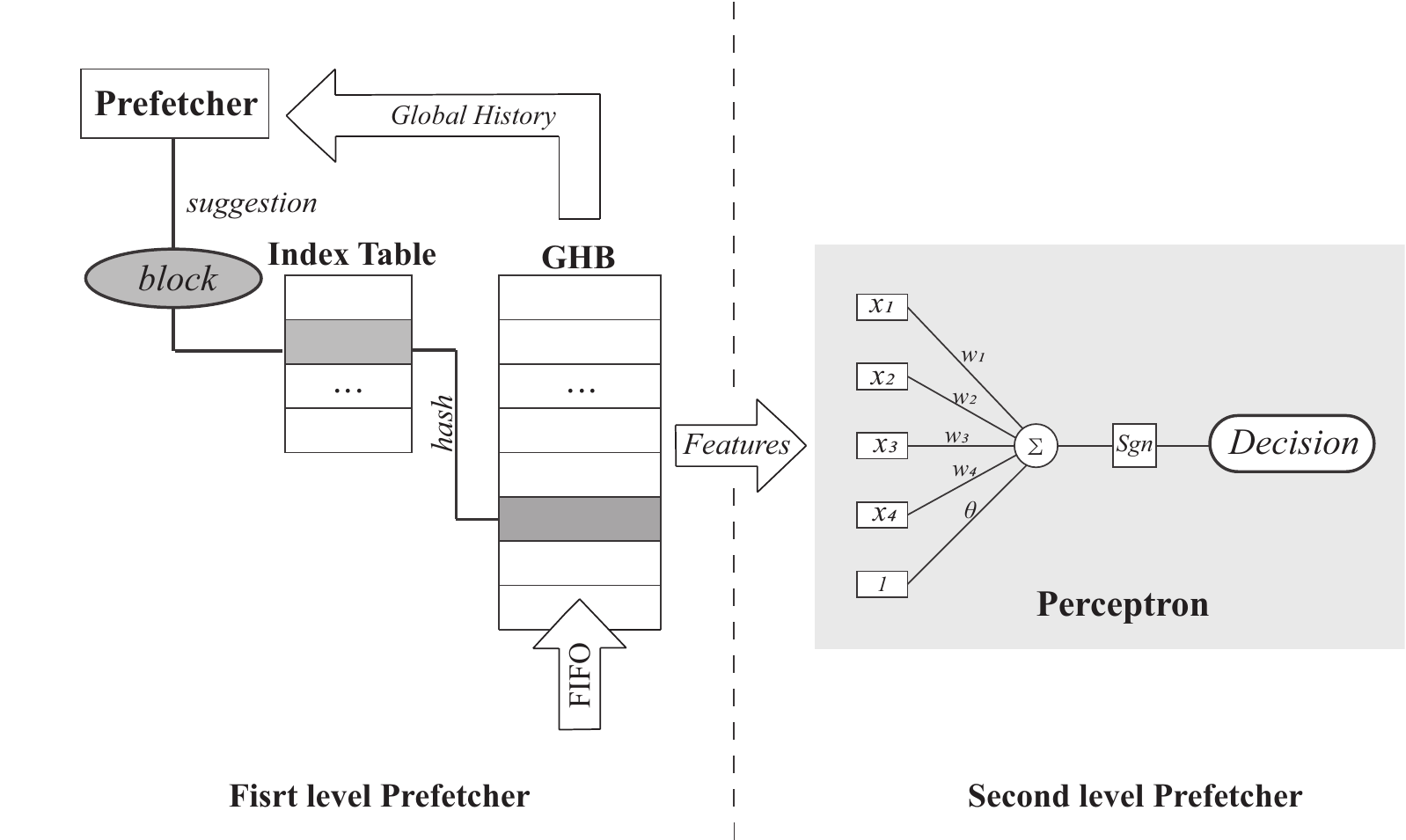}
			\caption{Two Level Prefetcher}
			\label{Fig:two-levelPrefetcher}
		\end{figure}
	\end{center}
	
	\subsection{Prefetching with Perceptron Learning}
	
	In this paper, we propose a two-level prefetcher, shown in Figure~\ref{Fig:two-levelPrefetcher}. The main idea is equipping the previous table-based prefetcher with the ability of learning by using artificial network.
	The first level is the combination of previous table-based prefetcher with the GHB, providing primary suggestion about which cache line to prefetch, and meanwhile, related information about the cache line are given to the next level. And in the second level, the perceptron will make use of those quantitative information to determine whether the prefetching request should be send to the cache read queue. This structure fully exploit local and global, time and space historical information provided by each cache miss.

	\begin{center}
		\begin{figure}[ht]
			\hspace{0em}
			\includegraphics[scale=0.73]{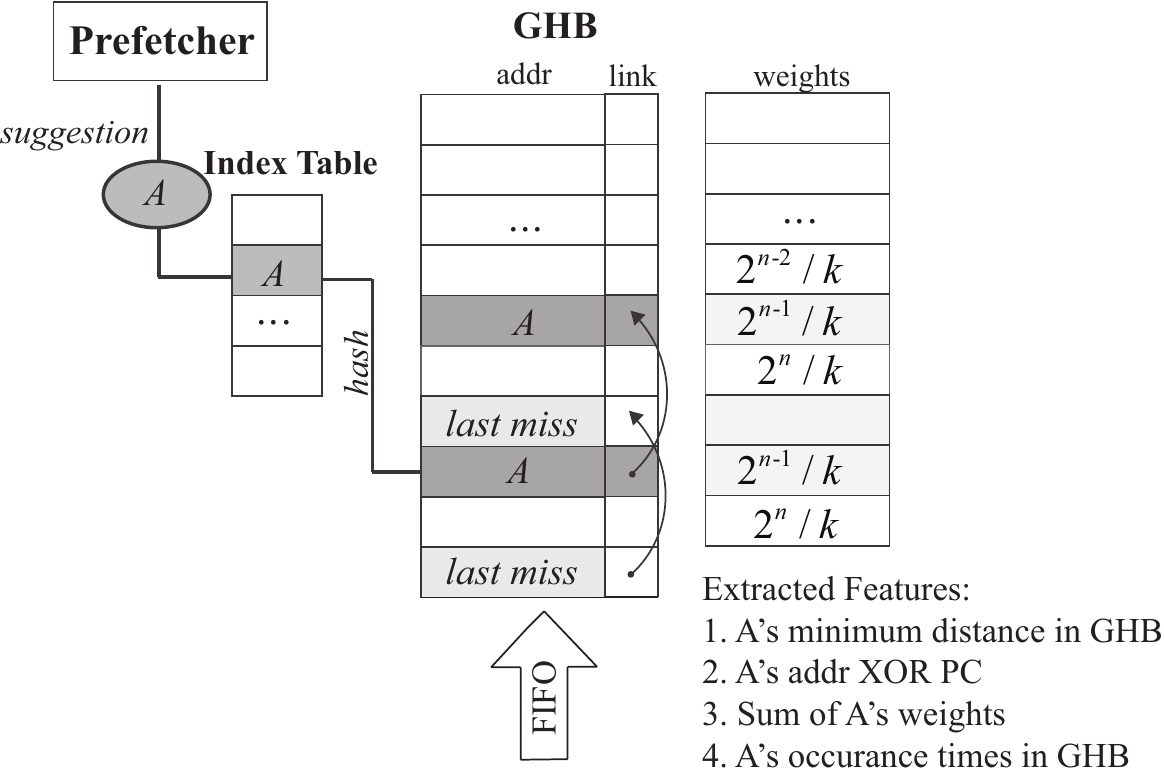}
			\caption{Memory access features extraction from the GHB and quantification}
			\label{Fig:features}
		\end{figure}
	\end{center} 
	
	Multiple linearly independent quantified features are correlated with prefetching determination consist the input vector for perceptron, as shown in Figure~\ref{Fig:features}. The features contain:

	\begin{enumerate}
		\item \textit{Prefetch distance: }
		According to the definition, prefetch distance indicates how far ahead of the demand access stream, the data block are prefetched\cite{refSurvey}. If the block has high prefetch distance, it shows that the block occupy cache space for a long time without being used by the processor, which is a kind of cache pollution. So this index represent the urgency of the block. Usually, this index uses CPU cycles as dimension. However, it costs high to get accuracy value for we have to trace every cache block. Instead, we use the block's minimum distance in the GHB as an alternative. Noticing that items of cache miss address in the GHB are stored according to their time sequential, this distance can be regarded as approximation time interval of processor require the same block. The shorter the distance, the more urgent that CPU needs this block. This index provide a local time information for the perceptron.
		\item \textit{Transition probability: }
		When a cache miss happen, a new entry will be pushed into GHB. The entries near to the last miss cache line has a high possibility of being used in future memory access. The neared, the more possible.  We can use transition probability, the term used in Markov diagram, to quantify this feature. For example, in Figure~\ref{Fig:features} the last miss cache line is found $k$ times of appearance. Then, along GHB time sequence, each entries is given a weight of $2^{n-m}/k$, in which $m$ indicates the distance between this entry and last miss cache line.
		The prefetcher gives the suggestion of prefetching $A$ into the cache, then all the weights relevant to $A$ are summed as the presentation of transition probability.
		\item \textit{Block address bits XOR program counter: }
		Combining program counter bits and block address bits by XOR indicates address distance delta. Due to the spatial and temporal principle of program, this index shows both time and space distance of the block.
		\item \textit{Occurrence frequency: }
		Noticing that GHB stores the most recent cache miss address, if a block frequently appears in GHB, it means that the processors requires the block several times, but it was wrongly replaced from the cache. So, the higher the value is, the higher the possibility that the block should be prefetched.
		\item \textit{Fixed input -- a constant: }
		This is a virtual feature. A threshold is needed by perceptron to make judgment. If the output  exceeds some threshold, the block will be determined as to be prefetched, and vise versa. However, it is impossible to find a fixed threshold suitable for all. Thus, to achieve the best performance, different threshold should be given for different workloads. Our solution is to give an additional input set to 1, and accordingly, a new weight. This new weight can works as the threshold and meanwhile, being adaptive to versatile workloads.
		
	\end{enumerate}

	In this way, our perceptron combines multiply features relating to program memory access pattern. To make a decision, the sum of quantitative feature and weight product are calculated as $y_{out}$ . If $y_{out}$ exceeds $0$, then the block are decided to be prefetched. And on the contrary, the block are decided as not to be prefetched.
	
	We use error-correction learning rule for perceptron training\cite{refError}. The weights are updated as 
	$$w_{j}\left ( t+1 \right )= w_{j}\left ( t \right ) + \alpha \left ( d - r \right ) x_{j}$$
	
	where $x_{j}$ indicates each input, $w_{j}$ is the according weight, $\alpha$ is the learning rate, $d$ is the desired value and $r$ is the real output of current status.
	
	\subsection{Prefetcher Organization}
	
	\begin{center}
		\begin{figure}[ht]
			\includegraphics[scale=0.5]{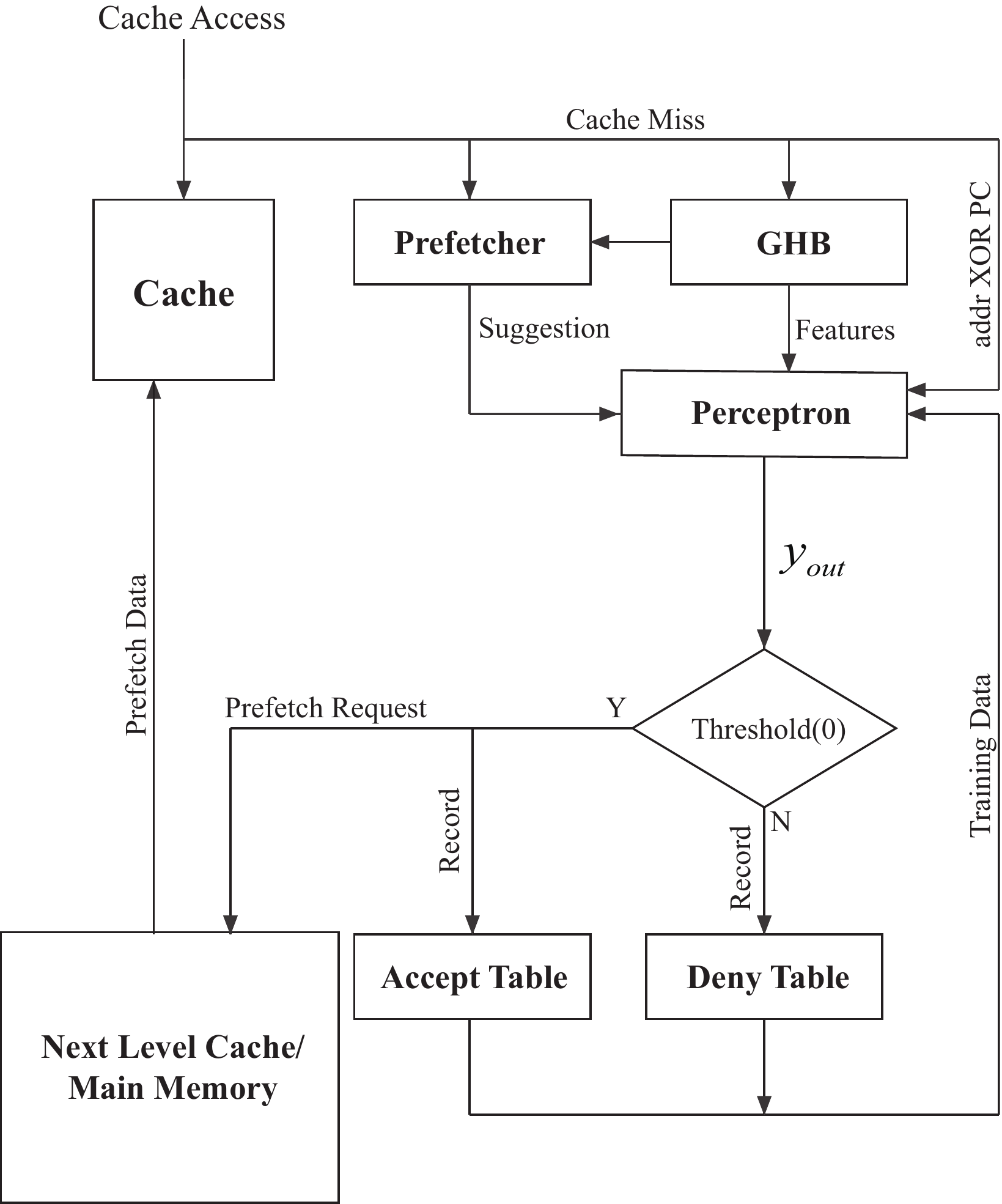}
			\caption{Datapath from extracting features from each cache miss and first level prefetcher giving suggestions, to perceptron making determinations, acting on it and receive training from feedback}
			\label{Fig:datapath}
		\end{figure}
	\end{center}
	
	The data path of our prefetching mechanism is shown in Figure~\ref{Fig:datapath}. Once a cache miss occurs, a new entry of the GHB will be pushed. In the mean time, the prefetcher will be triggered and gives suggestions of prefetching addresses. Those blocks are searched in the GHB, then the related information will be extracted in the GHB, and delivered to perceptron as input. After being quantified, the perceptron calculate the sum of each input and weight, as $y_{out}$. If $y_{out}$ exceeds $0$, this suggestion will be taken and the prefetch request will be sent to next level cache or main memory through cache read queue. Also a new record of this acceptance will be pushed into accept table. Otherwise the suggestion will be denied and recorded in deny table. both records in accept table and deny table will be further used as training data for perceptron in order to make adjustment to possible memory access pattern changes.
	
	\subsection{Hardware Implementation}
	With limited timeliness in processors, prefetcher should urgently conduct its action in several clock cycles. Our two-level prefetcher inevitably introduces extra latency. We propose a solution here. Multiple hardware-implemented perceptrons share one copy of weights are instantiated in prefetcher. The quantity of instances depends on first level prefetch degree. Thus perceptrons work simultaneously to give determinations in parallel. Perceptron would dynamically gain training and weights would be updated to adapt to program memory reference behavior changes. This scheme compresses the latency in a feasible cost of transistors.

	\section{Methodology}
	
	\subsection{Implementation}
	
	Prefetcher implementation is a systematic issue in which multiple factors should be considered. Since our simulation is intended to evaluate performance increase brought by perceptron learning, other factors such as prefetch trigger are not our major concerns as long as conditions are kept same in all experiments. We simply trigger prefetching after each cache miss.
	
	We implemented four prefetchers, including two conventional methods (Stride prefetcher and Markov prefetcher), and two two-level prefetchers (Stride prefetcher with perceptron, Markov prefetcher with perceptron). The prefetch degree of stride prefetcher is 2, and of Markov prefetcher is 4. Both of the four prefetchers are equipped with a 512 entry GHB.\cite{refGHB1,refGHB2}
	
	For each prefetch decision, we extract features from GHB and Program Counter, send data to perceptron. If the perceptron supports the suggestion from first level prefetcher, the request will be sent to the next level interconnect. Meanwhile the decision made by perceptron will be recorded in accept table or deny table.
	
	Our scheme needs supervised on-line learning, namely, feedback being required for further perceptron training. The best way is to trace each cache line access in a long term. However, this is infeasible in both time and space. So we use cache access interval as an alternative.
	
	For items in accept table, if haven't been accessed within 256 cache reference, are being treated as wrong prediction. And for items in deny table, if haven't been accessed within 32 cache misses (32 prefetch triggers), are being treated as a correct deny. Both these information will be used as feedback for perceptron on-line training.

	\subsection{Spatial Overhead}
	
	Our simulation works for x86-64 architecture\cite{refHaswell,refIntel}, which has 48-bit virtual address and 64-bit cache line size. That means we can record a cache line address in 45 bits.
	
	GHB-based stride prefetcher and GHB-based Markov prefetcher requires a global history buffer to record history cache miss data. Since we choose a 512 entry GHB and a 64 bits cache line size, each GHB entry use a 45 bits files to record cache line address, and a 9 bits fields to record links in GHB. Stride prefetcher does not need additional space while Markov prefetcher need an extra index table. Both prefetcher requires 3.375KB in the GHB, and Markov prefetcher consume approximately 1KB additional space for index table.
	
	Each entry in accept table and deny table requires a 45 bits field to record prefetching cache line address, a 32 bits fields to save perceptron inputs, and a 8 bits field to save duration. Each cycle this field increase by 1, and this cache line would be treated as not used once this 8-bits counter overflow or a new entry enters. In our simulation we use a 256 entry accept table and a 32 entry deny table. Thus, the perceptron decision trace consumes 3KB.
	
	Perceptron requires 5 8-bits registers to save weights, which bring minor spatial overhead.
	
	Overall overhead of each prefetcher is listed in Table~\ref{table:overhead}.
	
	\begin{scriptsize}
		\begin{table}[h!]
			\centering
			\begin{tabular}{|p{1.8cm}|p{6cm}|}
				\hline
				\textbf{Technique} & \textbf{Overhead}\\
				\hline
				\hline
				Stride & 3.375KB for GHB = 3.375KB \\
				\hline
				Stride + & 3.375KB for GHB + 3KB for perceptron \\
				Perceptron & decision trace = 6.375KB \\
				\hline
				Markov & 4.375KB for GHB and Index Table = 4.375KB \\
				\hline
				Markov + & 4.375KB for GHB and Index Table + 3KB \\
				Perceptron & for perceptron decision trace = 7.375KB \\
				\hline
			\end{tabular}
			\caption{Spatial overhead required by the various techniques is approximately the same in our implementation}
			\label{table:overhead}
		\end{table}
	\end{scriptsize}
	
	\subsection{Simulation}
	
	In our work, we made fully simulation to evaluate our two-level prefetcher scheme. We model performance with MARSSx86\cite{refMarss} simulator, configured using default Intel Xeon single core\cite{refHaswell,refIntel}. It models an out-of-order execution pipeline with parameters listed in Table~\ref{table:CPUConfig}. Cache configuration is listed in Table~\ref{table:cacheConfig}.

	\begin{scriptsize}
		\begin{table}[h!]
			\centering
			\begin{tabular}{|l|l|}
				\hline
				\textbf{Item} & \textbf{Configuration}\\
				\hline
				\hline
				Issue Width & 5 \\
				\hline
				Commit Width & 4 \\
				\hline
				ROB Size & 128 \\
				\hline
				ALU Units & 6 \\
				\hline
				FPU Units & 6 \\
				\hline
				Load Units & 1 \\
				\hline
				Store Units & 1 \\
				\hline
				Load Queue Size & 48 \\
				\hline
				Store Queue Size & 32 \\
				\hline
			\end{tabular}
			\caption{Pipeline Configuration in simulation using default Intel Xeon single core parameters}
			\label{table:CPUConfig}
		\end{table}
	\end{scriptsize}

	\begin{scriptsize}
		\begin{table}[h!]
			\centering
			\begin{tabular}{|l|l|l|l|l|}
				\hline
				\textbf{ } & \textbf{L1 D} & \textbf{L1 I} & \textbf{L2} & \textbf{L3}\\
				\hline
				\hline
				Type & WB & WB & WB & MESI Coherent \\
				\hline
				Size & 32KB & 32KB & 256KB & 12MB \\
				\hline
				Line Size & 64 & 64 & 64 & 64 \\
				\hline
				Associative & 8 & 4 & 8 & 16 \\
				\hline
				Latency & 4 & 2 & 6 & 27 \\
				\hline
				Read Ports & 2 & 2 & 2 & 2 \\
				\hline
				Write Ports & 1 & 1 & 2 & 2 \\
				\hline
				Prefetcher & None & None & {\bf Testing} & None \\
				\hline
			\end{tabular}
			\caption{Cache Configuration in simulation using default Intel Xeon single core parameters}
			\label{table:cacheConfig}
		\end{table}
	\end{scriptsize}

	We Implement our two-level cache on L2 cache with the following reasons:
	\begin{enumerate}
		\item Our scheme intents to decrease amount of memory requests issued by prefetcher, and would bring additional latency as a trade-off. Thus it's not suggested for latency intensive caches such as L1 cache.
		
		\item L2 cache is the last level cache for many processors.
		
		\item Each cache level filters a part of access requests. If L3 cache exists and our prefetcher work in L3 cache, enough features cannot be ensured for perceptron to make reasonable determination with only minor part of information left.
		
		\item For Intel Xeon processor we simulated, L3 cache is a coherent cache, which means we cannot extract the relationship between cache misses and PCs. This additional complexity would bring interference to our implementation.
	\end{enumerate}
	
	For memory we use default configuration in MARSSx86, a single channel main memory with 54ns access latency.
	
	\subsection{Workloads}
	
	We use the 29 SPEC CPU 2006 benchmarks\cite{refSPEC}. Each benchmark is compiled using default configuration file distributed with SPEC CPU 2006 benchmark. we use ptlsim integrated in MARSSx86 to create and restore from checkpoint. For each benchmark, the first 20 million instructions are used to warm micro-architectural structures, then the subsequent 1 billion are used to generate report results. For each benchmark we run 3 times to avoid accidental error.

	\section{Results}
	
	This section gives the performance evaluation results of stride prefetch, Markov prefetch and their perceptron variant. It presents IPC, cache misses, prefetch fails, perceptron accept suggestions, and prefetch counts. To simplify the discussion and illustrate the relationship between method, the names given to each prefetching methods are as follows: stride prefetcher: S, stride prefetcher with perceptron learning: SP, Markov prefetcher: M, and Markov prefetcher with perceptron learning: MP.
	
	\subsection{IPC}
	
	\begin{figure}[h!]
		\begin{flushleft}
			\includegraphics[scale=0.51]{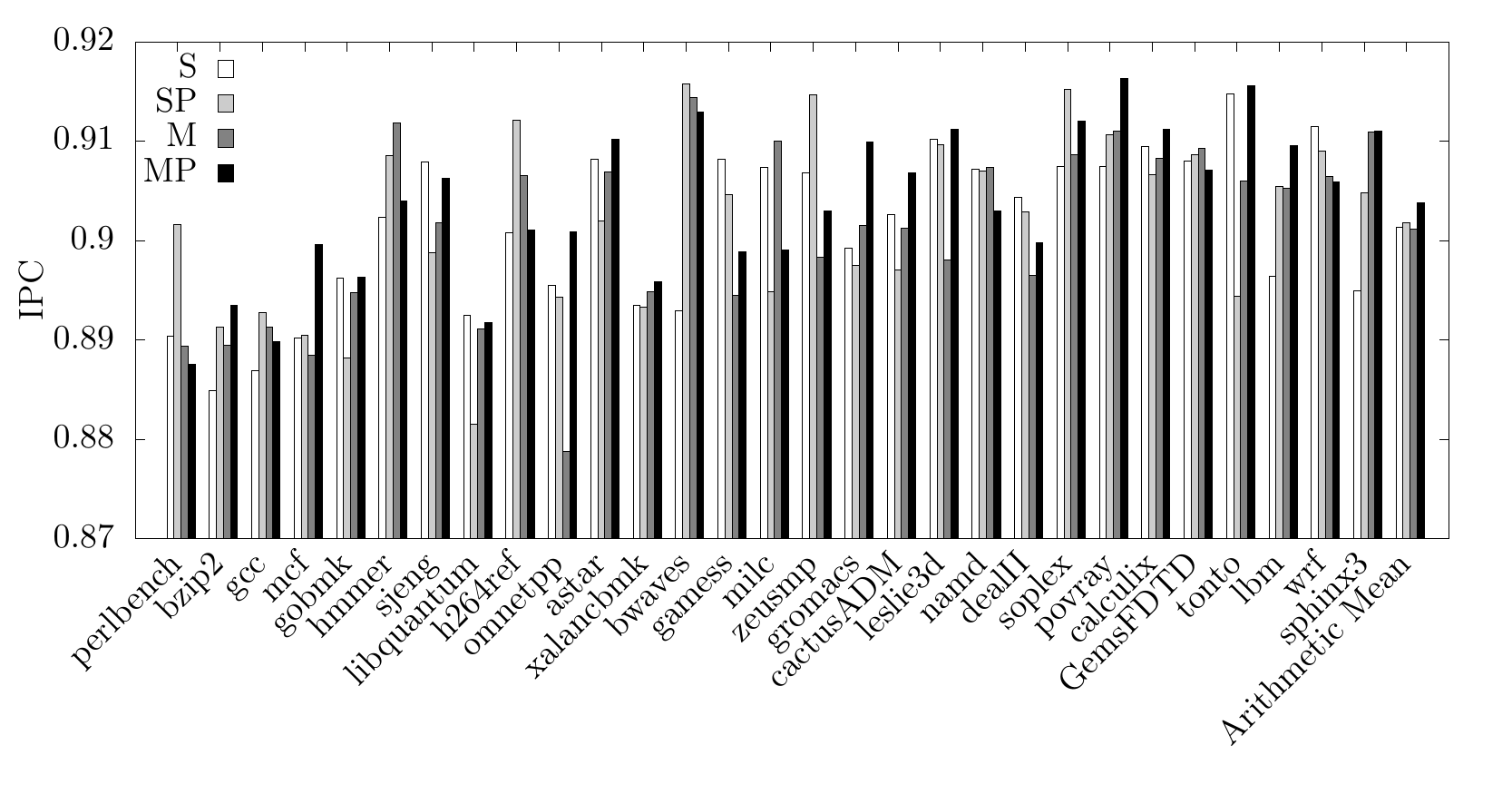}
		\end{flushleft}
		\caption{IPC for single core workloads}
		\label{Fig:IPC}
	\end{figure}
	
	\begin{figure}[h!]
		\begin{flushleft}
			\includegraphics[scale=0.51]{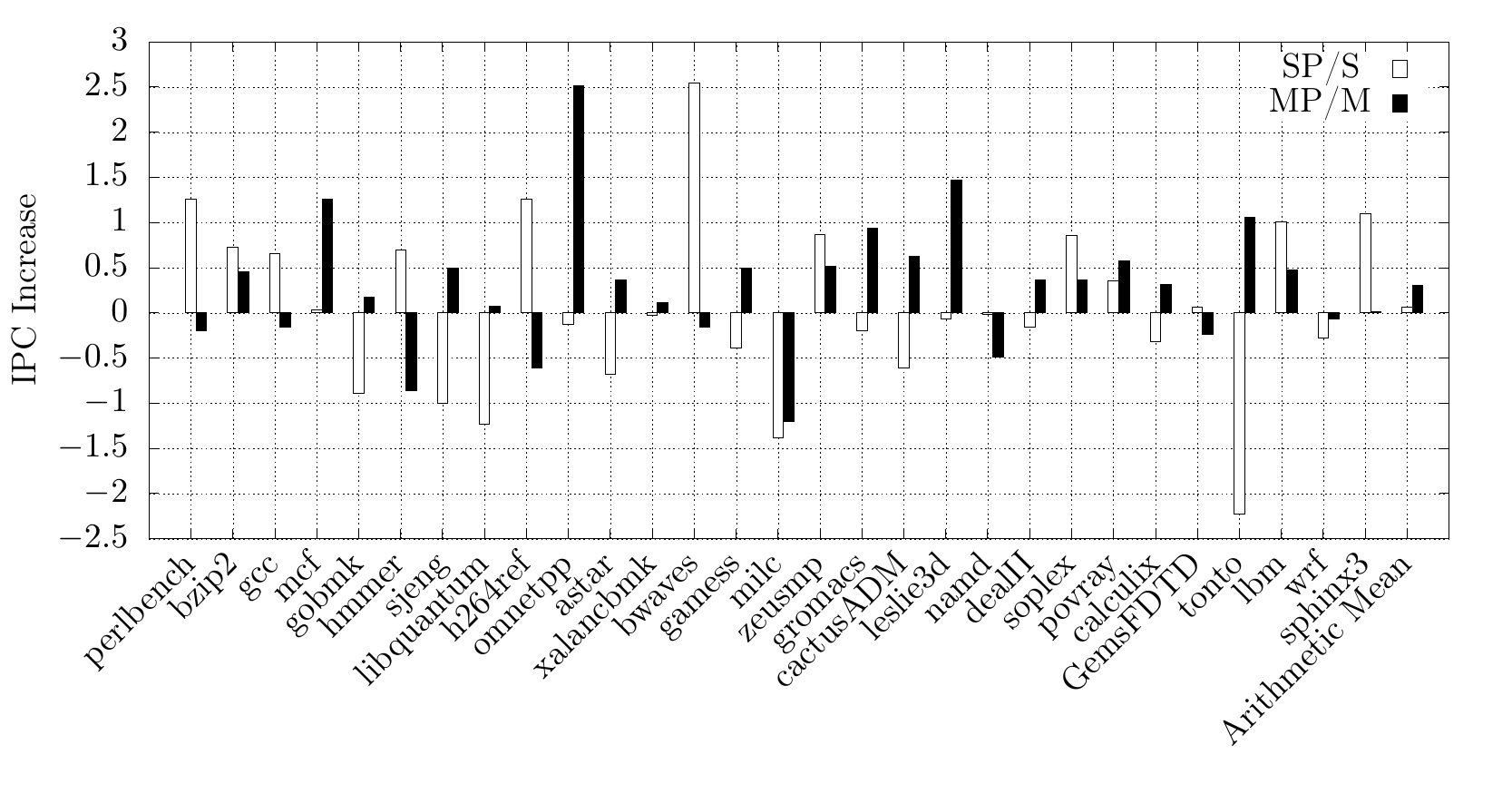}
		\end{flushleft}
		\caption{IPC for single core workloads increase SP achieved over S and MP  achieved over M}
		\label{Fig:IPCInc}
	\end{figure}

	Figure~\ref{Fig:IPC} shows the IPC of each prefetcher in SPEC CPU 2006 benchmarks. And Figure~\ref{Fig:IPCInc} shows the IPC increase brought by perceptron. Data shows that perceptron learning does not exert negative influence on IPC (floating between -2.22\% and 2.55\%). We can see that with the help of perceptron learning, stride prefetcher achieves IPC boost for more than 1\% in perlbench, h264ref, bwaves, lbm, and sphinx3, Markov prefetcher, in mcf, omnetpp, leslie3d and tonto. On average perceptron learning bring IPC increase for 0.05\% and 0.29\% respectively, which indicates that minor performance increase are achieved due to our new scheme.

	\subsection{Prefetch Correct and Error}
	
	\begin{figure}[h!]
		\begin{flushleft}
			\includegraphics[scale=0.51]{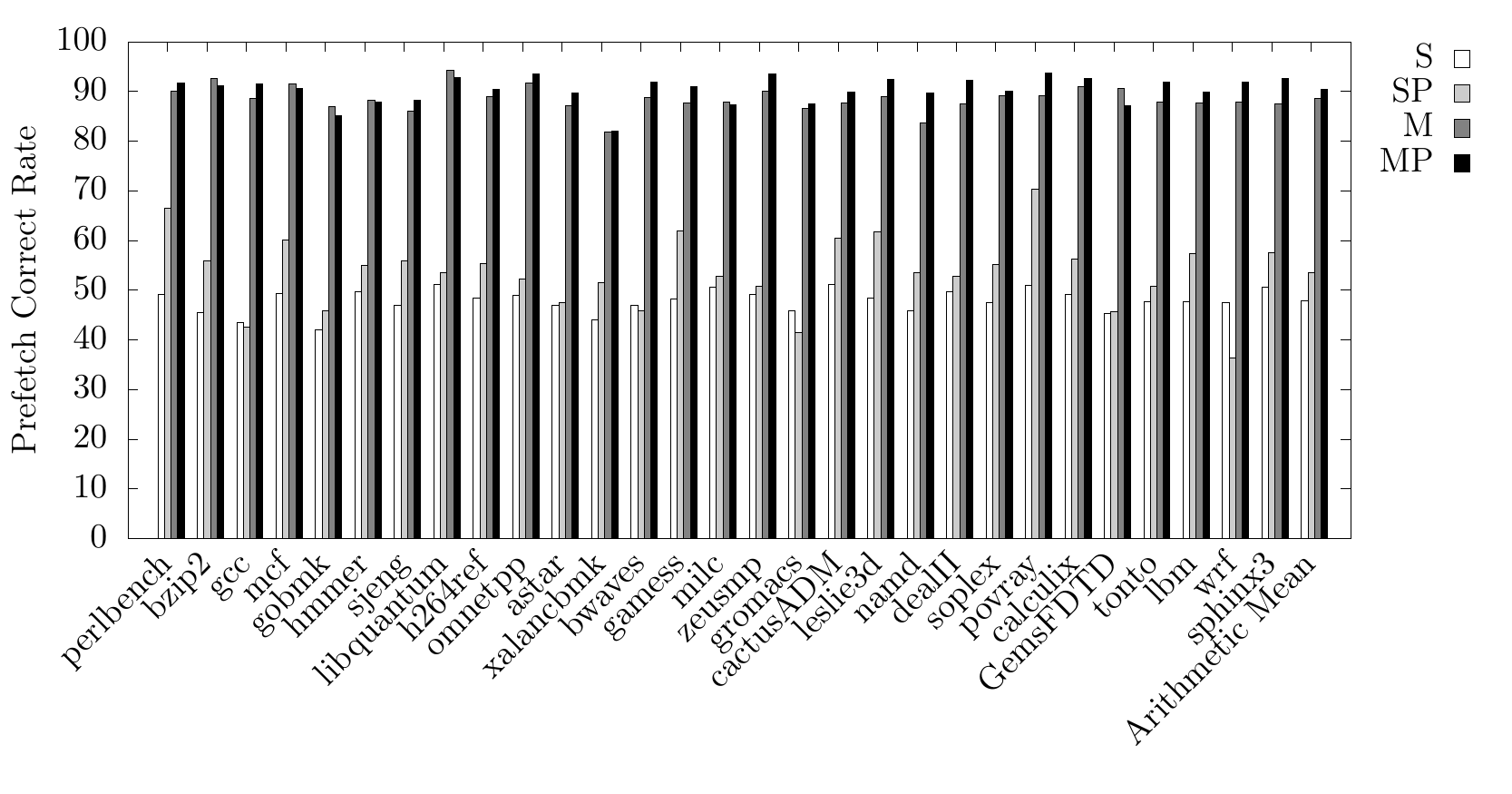}
		\end{flushleft}
		\caption{Prefetch Correct Rate in single core workloads}
		\label{Fig:CorrectRate}
	\end{figure}

	\begin{figure}[h!]
		\begin{flushleft}
			\includegraphics[scale=0.51]{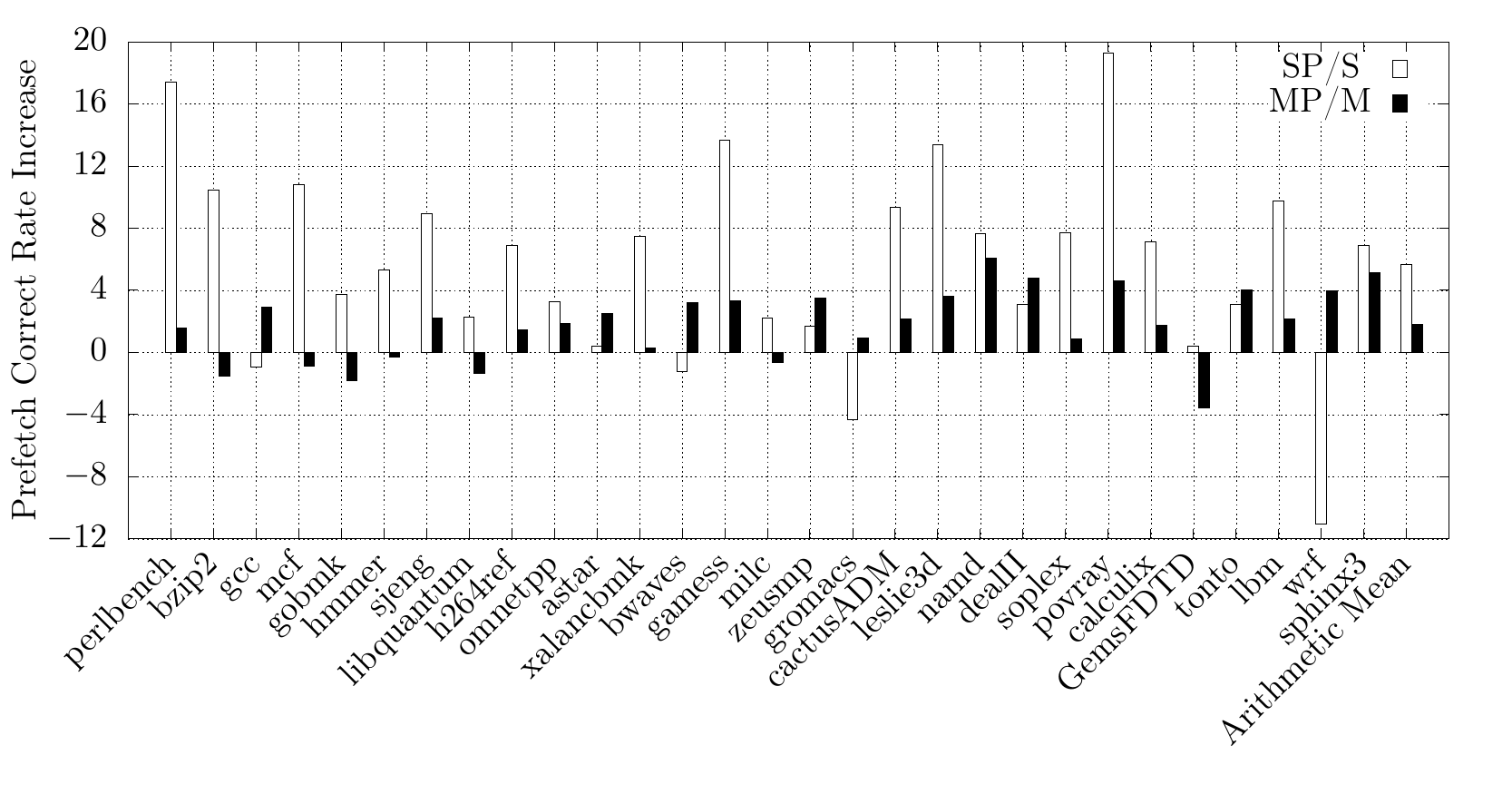}
		\end{flushleft}
		\caption{Prefetch Correct Rate in single core workloads increase SP achieved over S and MP achieved over M}
		\label{Fig:CorrectRateInc}
	\end{figure}
	
\begin{figure}[h!]
		\begin{flushleft}
			\includegraphics[scale=0.51]{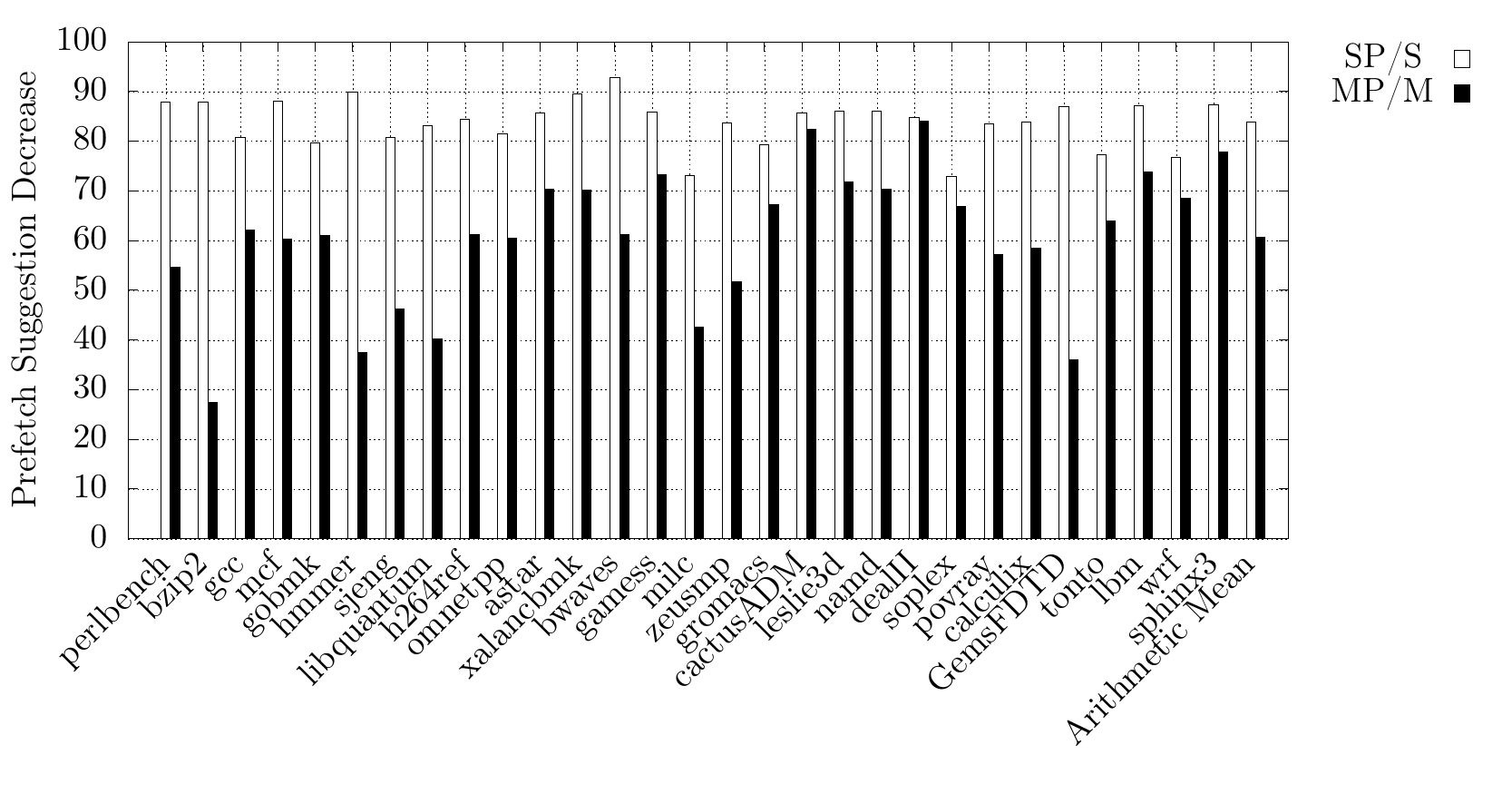}
		\end{flushleft}
		\caption{Prefetch Suggestion Decrease: the decrease of total suggestions issued by stride prefetcher and Markov prefetcher with perceptron learning to that without, respectively}
		\label{Fig:SuggestionDec}
	\end{figure}

	Figure~\ref{Fig:CorrectRate} and Figure~\ref{Fig:CorrectRateInc} shows a remarkable increase in prefetch correct rate rate in almost all SPEC CPU 2006 benchmarks. On average, compared with original prefetchers, our new scheme ameliorate prefetch correct rate by an arithmetic mean of 5.68\% and 1.82\%, respectively. Related with Figure~\ref{Fig:DenyRate} and Figure~\ref{Fig:SuggestionDec}, prefetcher now tends to make conservative decisions, saving any memory bandwidth that can be saved.
	
	\begin{figure}[h!]
		\begin{flushleft}
			\includegraphics[scale=0.51]{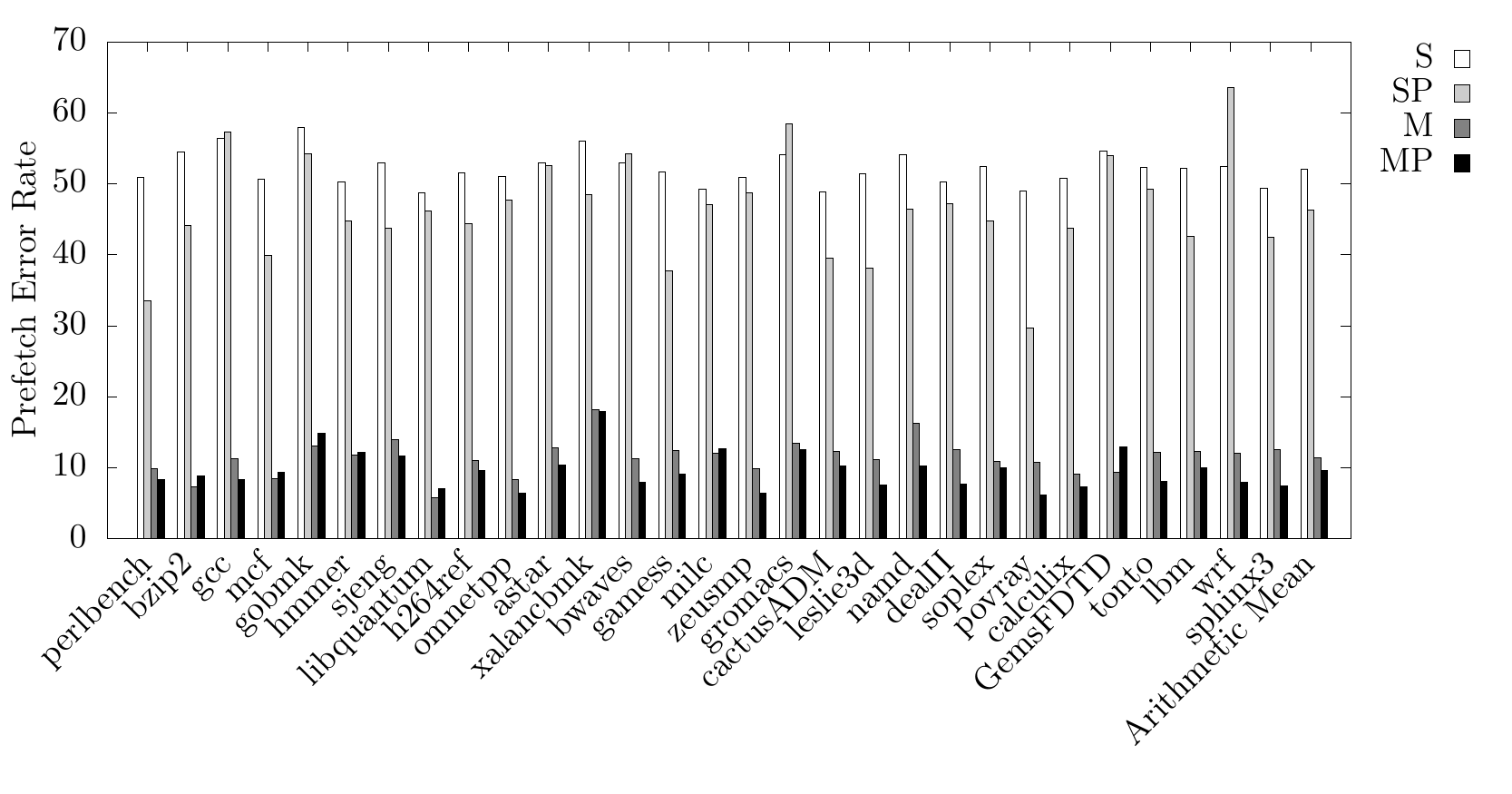}
		\end{flushleft}
		\caption{Prefetch Error Rate in single core workloads}
		\label{Fig:ErrorRate}
	\end{figure}
	
	\begin{figure}[h!]
		\begin{flushleft}
			\includegraphics[scale=0.51]{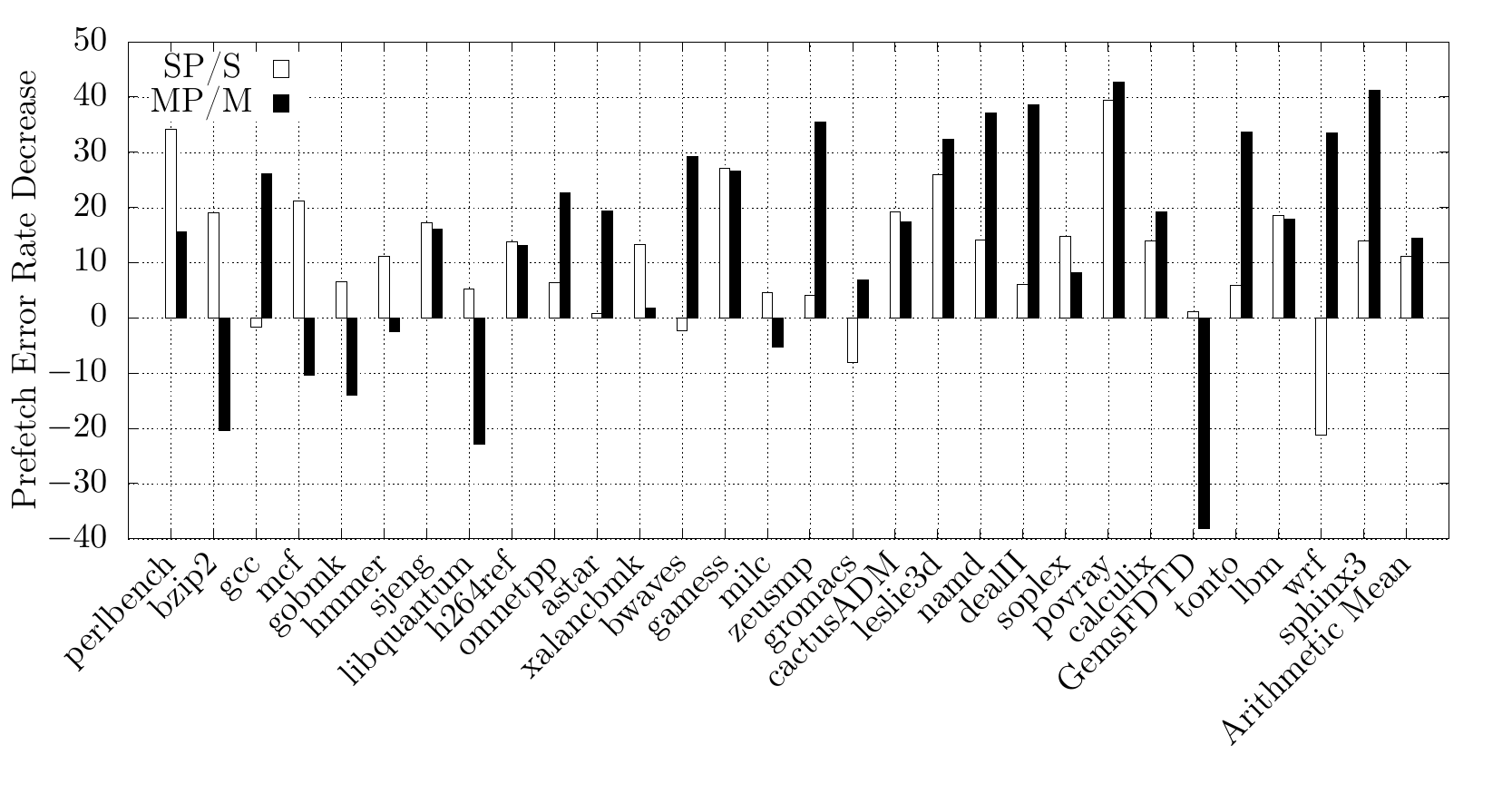}
		\end{flushleft}
		\caption{Prefetch Error Rate in single core workloads decrease SP achieved over S and MP achieved over M}
		\label{Fig:ErrorRateDec}
	\end{figure}
	
	Figure~\ref{Fig:ErrorRate} and Figure~\ref{Fig:ErrorRateDec} shows a significant decrease in prefetch error rate in most SPEC CPU 2006 benchmarks. On average, compared with original prefetchers, our new scheme mitigates prefetch error rate by an arithmetic mean of 11.17\% and 14.5\%, respectively.
	
	\subsection{Perceptron Deny}
	
	\begin{figure}[h!]
		\begin{flushleft}
			\includegraphics[scale=0.51]{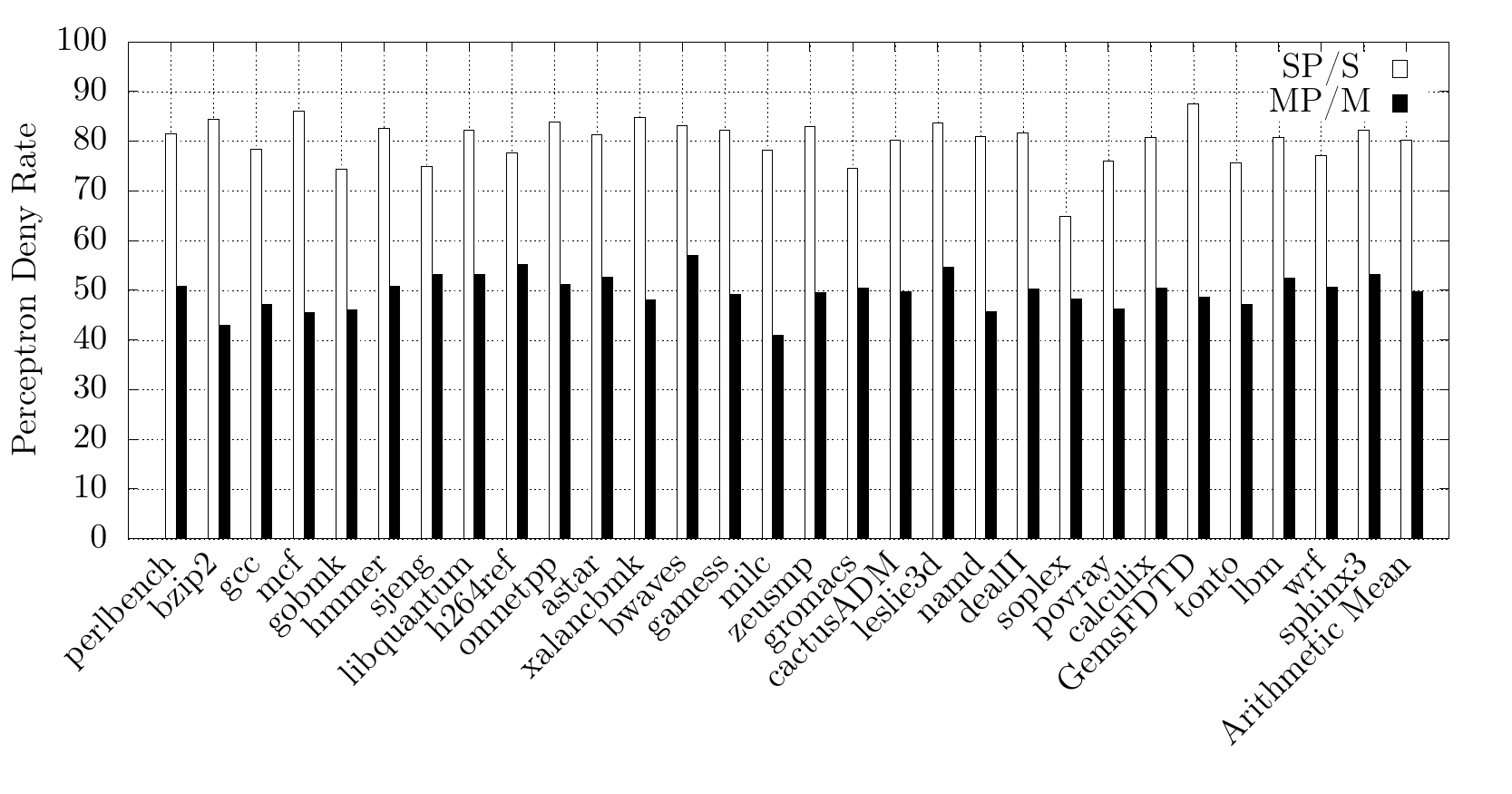}
		\end{flushleft}
		\caption{Perceptron Deny Rate: the ratio of the quantity of suggestions denied by perceptron to total suggestions issued by stride prefetcher and Markov prefetcher in the first level, respectively}
		\label{Fig:DenyRate}
	\end{figure}
	
	In Figure~\ref{Fig:DenyRate}, an arithmetic mean of 80.14\% of the unnecessary prefetcher suggestions issued by stride prefetcher and an arithmetic mean of 49.71\% of that issued by Markov prefetcher are detected and denied in the second level by perceptron.
	A large perceptron deny rate means a relatively lower memory request count, which would bring additional benefits to memory intensive workloads, and reduce memory bandwidth requirements, which alleviate memory pressure especially in multi-core system.
	
	\subsection{Prefetch Suggestion Count}
	
	Apart from the achievements, we have to mention that with perceptron learning, the suggestions from first level prefetching sharply decrease. We would like to give an explanation on this phenomenon.  Apart from program counter, the inputs of the perceptron are from GHB, which stores each cache misses. Perceptron's final decision directly exert influence on cache miss spatial distribution, which further changes items in GHB. We refer to this as a dynamic equilibrium. And our scheme would push this dynamic equilibrium to a new point of balance with the achievement of better performance in addressing cache pollution and cutting down memory traffic. We will make a detailed discussion in our conclusion.
	
	\subsection{Cache Hits}
	
	\begin{figure}[h!]
		\begin{flushleft}
			\includegraphics[scale=0.51]{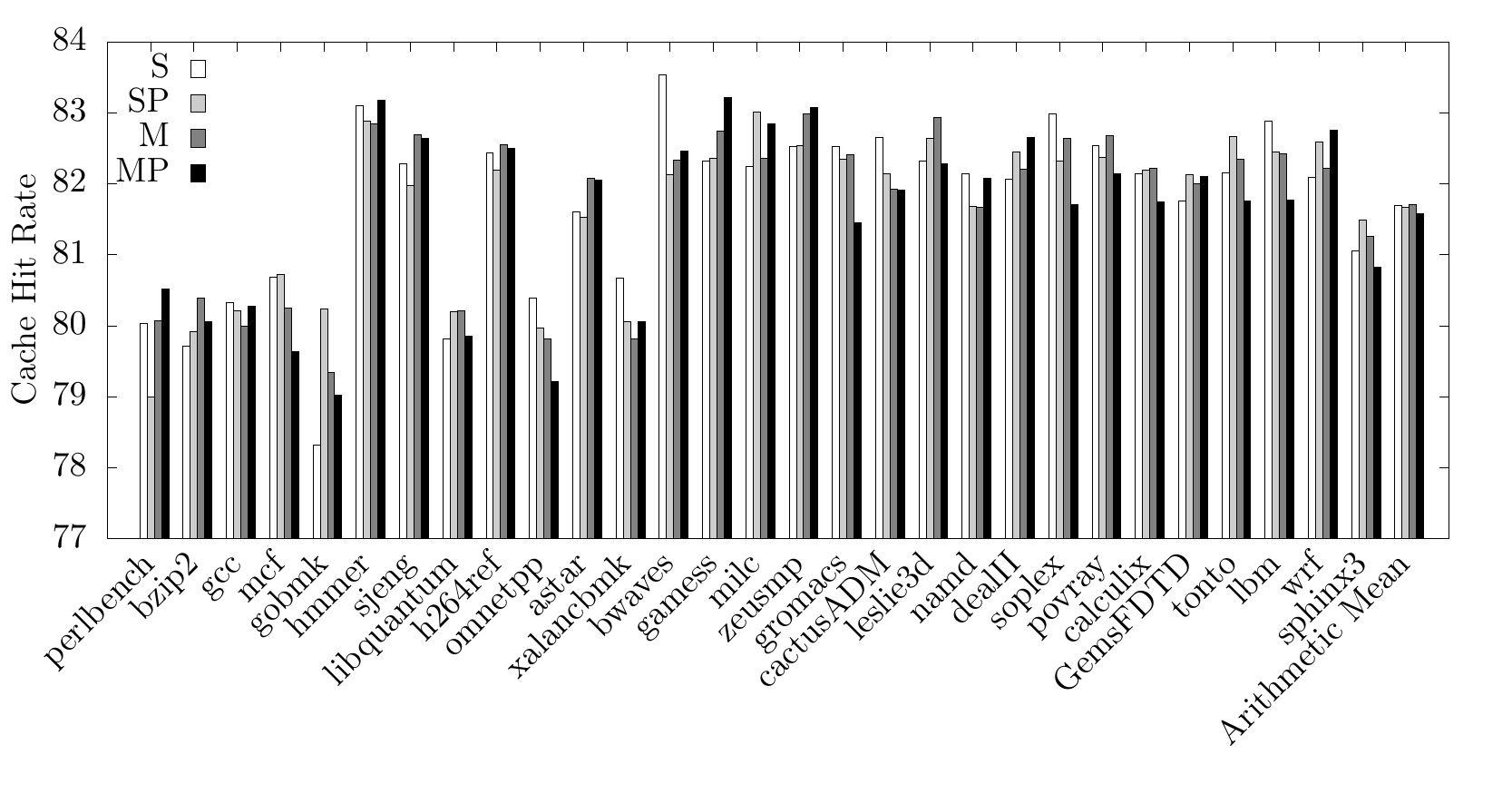}
		\end{flushleft}
		\caption{Cache Hit Rate in single core workloads}
		\label{Fig:HitRate}
	\end{figure}
	
	\begin{figure}[h!]
		\begin{flushleft}
			\includegraphics[scale=0.51]{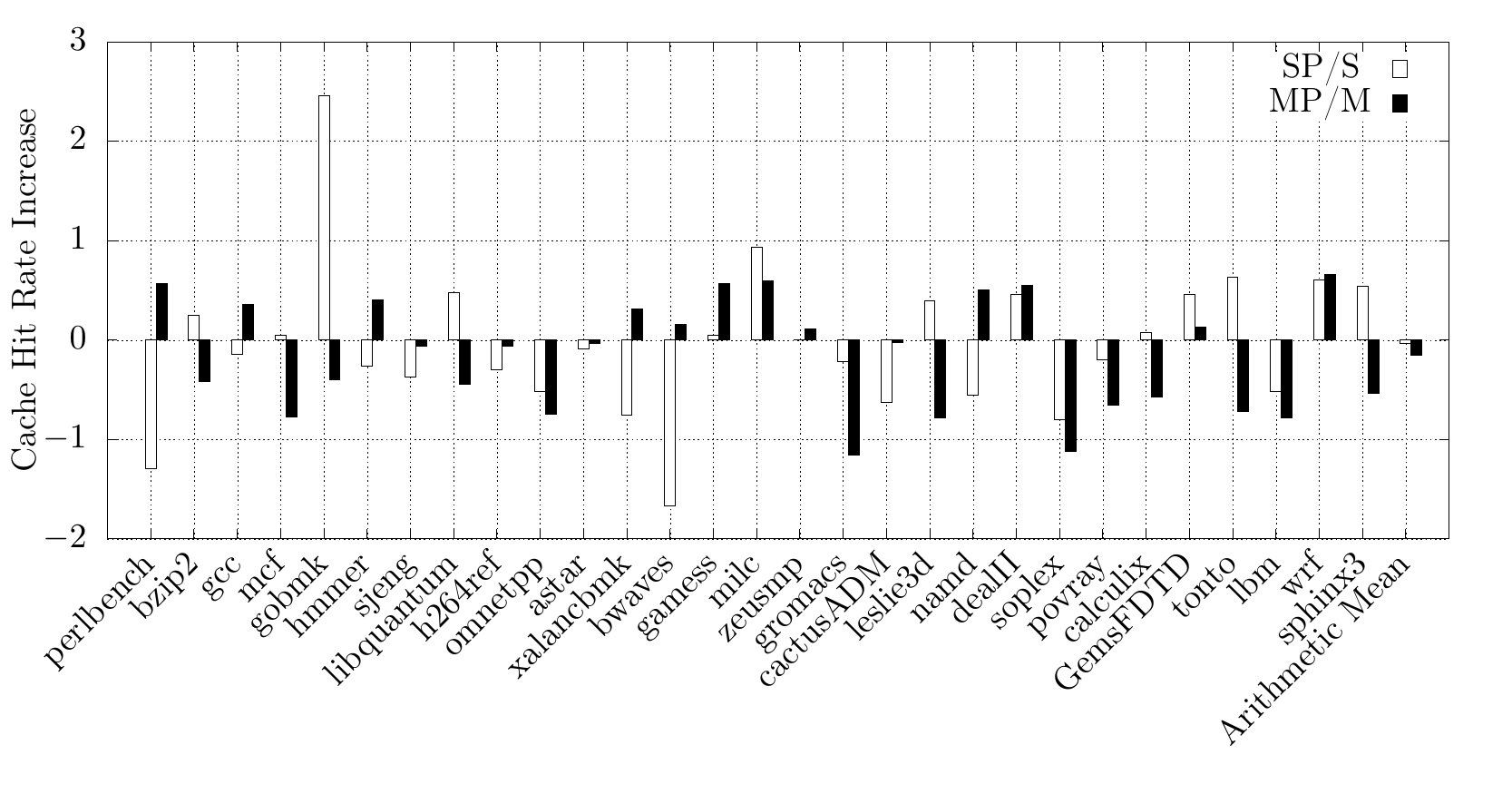}
		\end{flushleft}
		\caption{Cache Hit Rate in single core workloads increase SP achieved over S and MP achieved over M}\label{Fig:HitRateInc}
	\end{figure}
	
	Figure~\ref{Fig:HitRate} and Figure~\ref{Fig:HitRateInc} shows a minor floating of cache hit rate after using perceptron learning, on average, 0.03\% and 0.15\% decrease respectively. However, IPC improvements seemingly contrary to this. We suggest that lesser pressure on memory subsystem lead to this.
	
	While voting for the suggestions, perceptron inevitably deny some necessary prefetch requests, which further leads to cache misses. However, compared with significant mitigate in cache pollution, minor cache hit rate degreed is tolerable. As mentioned above, with perceptron learning the quantity of prefetch suggestions comes from first level prefetcher decrease greatly, less information is provided for perceptron to make decision. Thus, this minor decrease in cache hit rate would be reasonable.
	
	\section{Conclusion and Future Work}
	
	This paper proposes a new scheme of two-level prefetching, with previous table-based prefetching works in the first level for giving suggestions along with providing necessary related information and perceptron learning works in the second level for making final decisions. Rather than using fixed pattern, perceptron learning, with local and global, time and space history, can dynamically detect and trace program's memory access pattern. What's more, this scheme is easily implemented by hardware, which does not 'steal' cycles from the execution of instruction stream.
	
	Our simulation shows that perceptron denies a large quantity of unnecessary memory request and thus, ameliorating cache pollution and mitigating memory traffic, while exert minor influence on IPC and cache hit rate.

	\begin{center}
		\begin{figure}[ht]
			\hspace{2em}
			\includegraphics[scale=0.5]{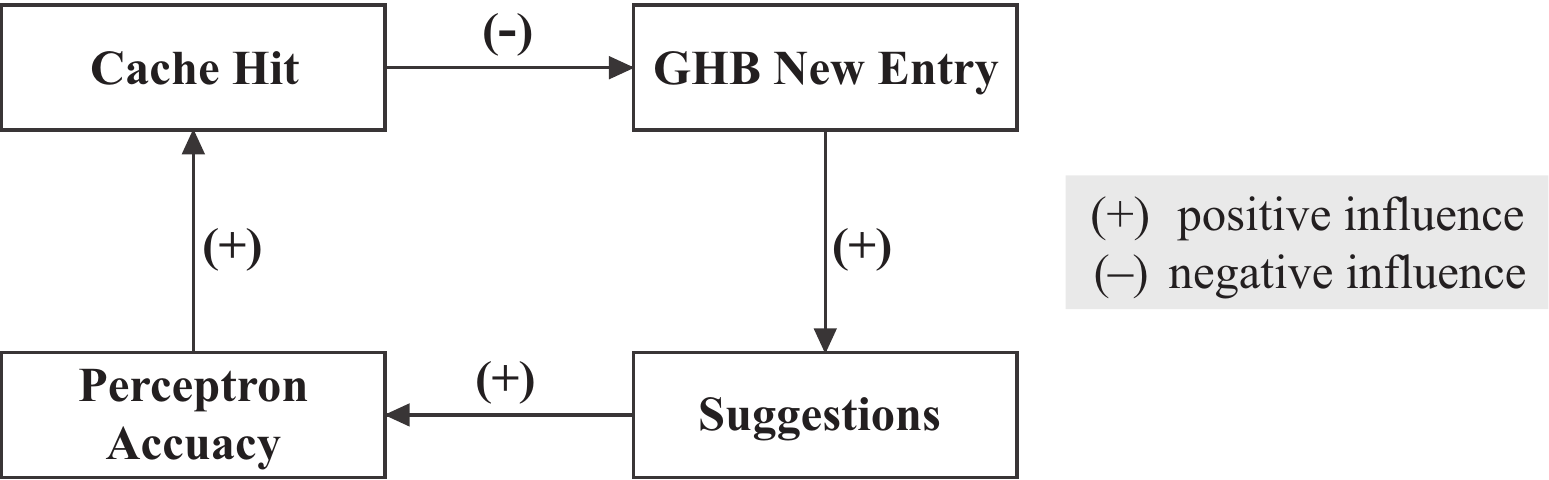}
			\caption{The Feedback Loop of Cache Hit}
			\label{Fig:loop}
		\end{figure}
	\end{center}
	
	Some furthermore issues should be discussed here. Simulation shows a sharp decrease in first level prefetcher suggestions. As shown in Figure~\ref{Fig:loop}, our interpretation is that a feedback loop exists in cache hit and first level prefetcher suggestions. A higher cache hit rate results in infrequent cache misses, thus less new entries are pushed in GHB. Consequently, GHB updates less frequently. Meanwhile, inadequate real time information leads to less suggestions issued by first level prefetcher. This may detriment perceptron in making reasonable determination due to inferior panorama of memory reference records.  Then decrease in perceptron accuracy further exert negative influence on cache hit rate. We refer to this as a dynamic equilibrium.  And our scheme would push this dynamic equilibrium to a new point of balance with the achievement of better performance in addressing cache pollution and cutting down memory traffic.
	
	This paper makes the following contributions: 
	1) A novel two-level prefetcher are proposed, which can effectively reduce unnecessary memory requests issued by prefetcher. This is vital for multi-core systems; 
	2) It shows that perceptron learning can be used to exploit and trace program memory access pattern dynamically; 
	3) Perceptron learning combined with previous table-based mechanisms can achieve large performance increase; and 
	4) Simulation modeled for fully evaluation of our scheme in dimensions including IPC, memory requests quantities, prefetching accuracy, cache hit rate and etc. is taken. 
	In future work, we intend to explore what is the best opportunity to trigger prefetcher. We also plan to explore how to maintain enough information in GHB for perceptron to make reasonable decisions.

	
	
	%
	
	\bibliographystyle{IEEEtranS}
	\bibliography{ms}
	
	

\end{document}